\documentclass[sigconf]{acmart}

\usepackage{lineno}
\usepackage{booktabs} 
\usepackage{graphicx}
\usepackage[table,xcdraw]{xcolor}
\usepackage{gensymb}
\usepackage{hyperref}
\usepackage{booktabs}
\usepackage{graphicx}
\usepackage{multirow}
\usepackage{enumitem}
\usepackage{tabularx}
\usepackage{hyphenat}
\usepackage{fancyhdr}
\usepackage{subcaption}
\usepackage[absolute,overlay]{textpos}

\renewcommand\footnotetextcopyrightpermission[1]{}

\acmConference[ASE-NIER '26]{41st IEEE/ACM International Conference on Automated Software Engineering}{October 12--16, 2026}{Munich, Germany}

\begin{document}

\title{Agents That Teach: Towards Designing Incidental Learning Back into AI-Assisted Software Development}

\author{Rohit Mehra$^\dagger$, Samdyuti Suri$^\dagger$, Prithviraj K Tagadinamani$^\dagger$, Kapil Singi$^\dagger$, Vikrant Kaulgud$^\dagger$, Adam P. Burden*} 
\affiliation{ 
	\institution{$^\dagger$Accenture Labs, India
		*Accenture, USA}
	\country{}
}
\email{{{rohit.a.mehra, samdyuti.suri, p.k.tagadinamani, kapil.singi, vikrant.kaulgud, adam.p.burden}@accenture.com}}

\begin{abstract}

AI coding agents are rapidly reshaping how software is built, with developers increasingly delegating substantial coding tasks to autonomous agents in pursuit of higher productivity. While these gains are real, they come at the cost of incidental learning. Developers historically acquired informal knowledge through effortful problem-solving, and this has long shaped how software engineering expertise develops. However, with over-reliance on agentic coding, unpracticed skills could atrophy silently over time. As this learning pathway is short-circuited, developers risk silently accruing Knowledge Debt, a developer-level analogue of Technical Debt, where changes the agent executes that the developer cannot fully understand accrue over time. In this paper, we argue that incidental learning will not re-emerge on its own and must be consciously designed back into developer-agent interactions, and propose six design principles to guide such systems. We then present \textit{SHIELD}, a multi-agent system grounded in the notion of \textit{agents that teach}, that operationalizes these principles by leveraging the AI coding agent's own reasoning to surface contextual, out-of-band learning moments without disrupting developer flow. Through this work, we envision a path toward learning-aware development environments where productivity and learning are complementary, not competing.

\end{abstract}


\maketitle

\begin{textblock*}{20.5cm}(1.9cm,26.2cm) 
	Accepted for publication at 41st IEEE/ACM International Conference on Automated Software Engineering (ASE 2026 - NIER Track)
\end{textblock*}

\section{Introduction}\label{introduction}

AI coding agents have moved from experimental tools to core components of modern software engineering workflows. Tools such as Claude Code, GitHub Copilot, and Cursor are now used not only for code generation, but also for critical activities such as debugging, refactoring, code review, and documentation, increasingly through autonomous, multi-step workflows \cite{claudecode, githubcopilot, cursor}. According to industry reports, approximately 42\% of all committed code is AI-generated or AI-assisted, with this share projected to reach 65\% by 2027 \cite{sonar}. This shift has translated into measurable productivity gains for developers, with recent studies documenting improvements in task completion rates and per-task development velocity in AI-assisted settings \cite{cui2025effects, 10.1109/ICSE-SEIP66354.2025.00060}. Yet, despite these gains, a more subtle but consequential shift is underway in how developers engage with these agents, one that suggests that while problems are increasingly solved, understanding/learning may not always follow.

While developers benefit from increased productivity and speed, this shift reflects a recurring pattern in human-tool interaction. As humans increasingly offload cognitive tasks to tools, the underlying capability tends to diminish over time. For example, sustained reliance on GPS has been linked to decline in spatial memory and navigation ability, spell check and autocorrect to weakened spelling ability, and smartphones to reduced memory recall ability \cite{dahmani2020gps, Ali_Nakshbandi_Saadi_Barzani_2024, 10.3389/fpubh.2024.1332030}. In each case, the tool absorbs the effort, and the human reduces active engagement with the underlying skill, leading to its gradual decline. AI coding agents appear to be following a similar trajectory, with emerging evidence suggesting erosion of core software engineering capabilities such as conceptual understanding, code reading, and debugging \cite{shen2026aiimpactsskillformation, chang2025codingaireflectionindustrial}. This reflects a broader pattern of \textit{skill atrophy}, where unpracticed capabilities do not simply remain dormant, but gradually deteriorate.

Before AI coding agents or assistants, a developer encountering a problem in code would search for answers, read through explanations, weigh alternative implementations, and ultimately manually write a fix they could understand and justify. The process was often slow and exploratory in nature, but along the way, the developer absorbed surrounding concepts such as unfamiliar APIs, architectural choices, alternatives, and trade-offs, gradually building expertise and evolving as a developer. Today, the same developer pastes the error into an agent and receives a working fix in seconds, often without understanding how or why it works. While the problem is solved, the learning associated with the resolution does not occur. What is lost in this shift is not formal training, but \textit{incidental learning}: the unintended acquisition of knowledge that occurs as a byproduct of effortful problem-solving \cite{marsick1990informal, 10.1145/3490099.3511138, Tran-Le_Thomas_Stiffler_Nguyen_2026}. Early evidence already supports this concern. In a controlled study, developers who used AI assistance scored 17\% lower on a subsequent comprehension assessment than those who completed the same tasks without AI, and developers who fully delegated coding to the AI showed the steepest decline in skill formation \cite{shen2026aiimpactsskillformation}. The exploratory process that once enabled incidental learning has been short-circuited, and with it, the implicit pathway through which software engineering expertise has historically been built.

We posit that the shift towards agent-driven development is creating a new and underexamined form of skill erosion in software engineering, one that is invisible in the short term but consequential over time. Developers remain productive in the short term while their independent capability quietly lags behind, leaving them able to build with AI but increasingly unable to debug, adapt, or extend that work on their own \cite{chang2025codingaireflectionindustrial}. Building on prior work in cognitive offloading and skill atrophy, this accumulating gap can be understood as a form of \textit{Knowledge Debt}: a developer-level analogue to Technical Debt, incurred when an agent makes changes to the codebase that the developer does not fully understand \cite{gerlich2025aitools, macnamara2024ai, 6336722}. Like Technical Debt, Knowledge Debt compounds silently and surfaces only when independent capability is required, at which point repayment through relearning becomes costly. \textit{We argue that, in the agentic era, incidental learning will not re-emerge on its own; it must be consciously designed back into developer-agent interactions so that Knowledge Debt is continuously repaid rather than allowed to accumulate.}

Addressing this problem requires moving beyond productivity as the sole measure of AI-assisted development and treating learning as a first-class design concern. In this paper, we introduce an initial set of design principles that approaches for embedding incidental learning into developer-agent interactions should satisfy. We further present \textit{SHIELD} (Safeguarding Human Expertise and Incidental Learning in Software Development), a novel multi-agent instantiation of the design principles, grounded in the notion of \textit{agents that teach}, which leverages the agent’s own reasoning to integrate incidental learning into developer workflows through out-of-band channels that preserve and respect developer flow, allowing knowledge gaps to be addressed as they arise. As agents take on a growing share of software development work, they must also help developers continue to learn, and not merely consume what is generated for them. The remainder of the paper develops these ideas and outlines an initial research agenda for the software engineering community.
\section{Related Work}\label{related_work}

Concerns about skill atrophy and displaced learning under AI have surfaced across multiple domains, with diverse interventions proposed to address them. In aviation and medicine, structural interventions such as scenario-based simulation training, minimum unaided practice, and active engagement checkpoints have been proposed to counter AI-induced deskilling \cite{Ong2026}. In broader knowledge work, cognitive forcing functions have been designed to reduce overreliance on AI by requiring deliberate engagement before accepting AI outputs, and Socratic scaffolding has been used in AI-assisted writing to preserve users' critical thinking during AI use \cite{10.1145/3449287, hugenroth2026}. Across these domains, the common thread is restoring human engagement through explicit interventions layered onto the workflow.

More recently, researchers have begun to tackle the same concern within software engineering as well. Initial efforts have characterized the resulting consequences, framing the accumulating gap between AI-generated code and developer understanding as epistemic debt or comprehension debt \cite{sankaranarayanan2026, ahmad2026}. In parallel, others have leveraged cognitive engagement techniques as a mechanism for preserving learning, requiring learners to engage with the problem before AI-generated code is revealed or accepted, through approaches such as lead-and-reveal, incremental reflection, and metacognitive teach-back gates \cite{10.1145/3708359.3712104, Tran-Le_Thomas_Stiffler_Nguyen_2026, sankaranarayanan2026}.

While current approaches layer learning onto the workflow as a structured activity that fires on every interaction, none directly target incidental learning. We approach learning as a natural byproduct of the work itself, embedded through ambient out-of-band channels that preserve flow, surfaced only when intervention is genuinely warranted, calibrated to each developer's existing knowledge, and delivered through agents distinct from the AI coding agent itself, as models optimized for code generation may not be well-suited for teaching. Together, these shifts motivate the framework and system we develop in the rest of the paper.
\section{Designing Incidental Learning into Developer-Agent Interactions}\label{incidental_learning_injection}

Designing incidental learning back into agent-assisted development requires reintroducing learning moments that are often displaced by AI agents, without reverting to slower non-AI workflows or imposing formal learning instructions. Such moments include reasoning and knowledge about why a specific API was chosen or why a particular design decision was made. The goal is not to teach the developer using a structured curriculum, but to surface lightweight, contextual, and relevant learning opportunities that emerge as a natural byproduct of agent-assisted coding, allowing Knowledge Debt to be repaid as it accrues. To this end, we propose six early design principles that, in our view, any approach embedding incidental learning into human-agent interactions should satisfy, and invite the community to refine and extend them.

\begin{itemize}[leftmargin=2.7em]
		
		\item \textbf{Contextual:} Learning opportunities must be tightly tied to the specific code, API, design choice, or trade-off the developer and AI coding agent just engaged with, rather than presented as generic or pre-authored content. They should illuminate the unfamiliar element directly in front of the developer, not redirect attention towards a general concept.
		
		\item \textbf{Grounded:} Decisions about what learning opportunities to surface must be grounded in the agent's reasoning, the explanations and intermediate steps it produces while generating code, rather than from observable artifacts like code changes or developer actions alone. The agent's reasoning carries information that cannot be reconstructed from artifacts alone, such as why a particular design decision was made, what alternatives were considered, and where the agent was uncertain. These are precisely the signals that distinguish genuine teachable moments from routine activity.
		
		\item \textbf{Ambient:} Learning interventions should live within the developer’s day-to-day environment, such as the IDE, where they can be noticed and engaged with quickly without requiring a context switch. They should respect the developer’s flow and current priorities, surfacing through out-of-band channels such as asynchronous queues, feeds, or peripheral panels, rather than through blocking dialogs or interruptive prompts. The developer should choose when to engage, not the system.
		
		\item \textbf{Selective:} The system should not fire on every interaction. It should decide whether a moment genuinely qualifies as a learning opportunity for the developer and intervene only when it does. Surfacing too frequently trains developers to ignore the system, ultimately undermining the very learning it aims to support.
		
		\item \textbf{Adaptive:} Learning interventions should be calibrated to the developer's expertise, prior exposure to similar tasks or concepts, and demonstrated familiarity. What is a learning moment for a junior developer may be noise for a more experienced one, and the system should account for this difference.
		
		\item \textbf{Closed-Loop:} The system must verify whether the learning has actually been internalized, through lightweight probes or comprehension checks, rather than treating delivery as success. Outcomes from each intervention should be used to update the system’s understanding of what to surface, deprioritize, or revisit in future interactions.
		
\end{itemize}

Taken together, these design principles describe what it means for an AI coding agent to support developer learning rather than displace it. They do not specify a system. Instead, they define what any such system must do. In the next section, we present SHIELD, an early instantiation that operationalizes these design principles.
\section{SHIELD}\label{shield}

\begin{figure}[t]
	\centering
	\includegraphics[width=1.0\linewidth]{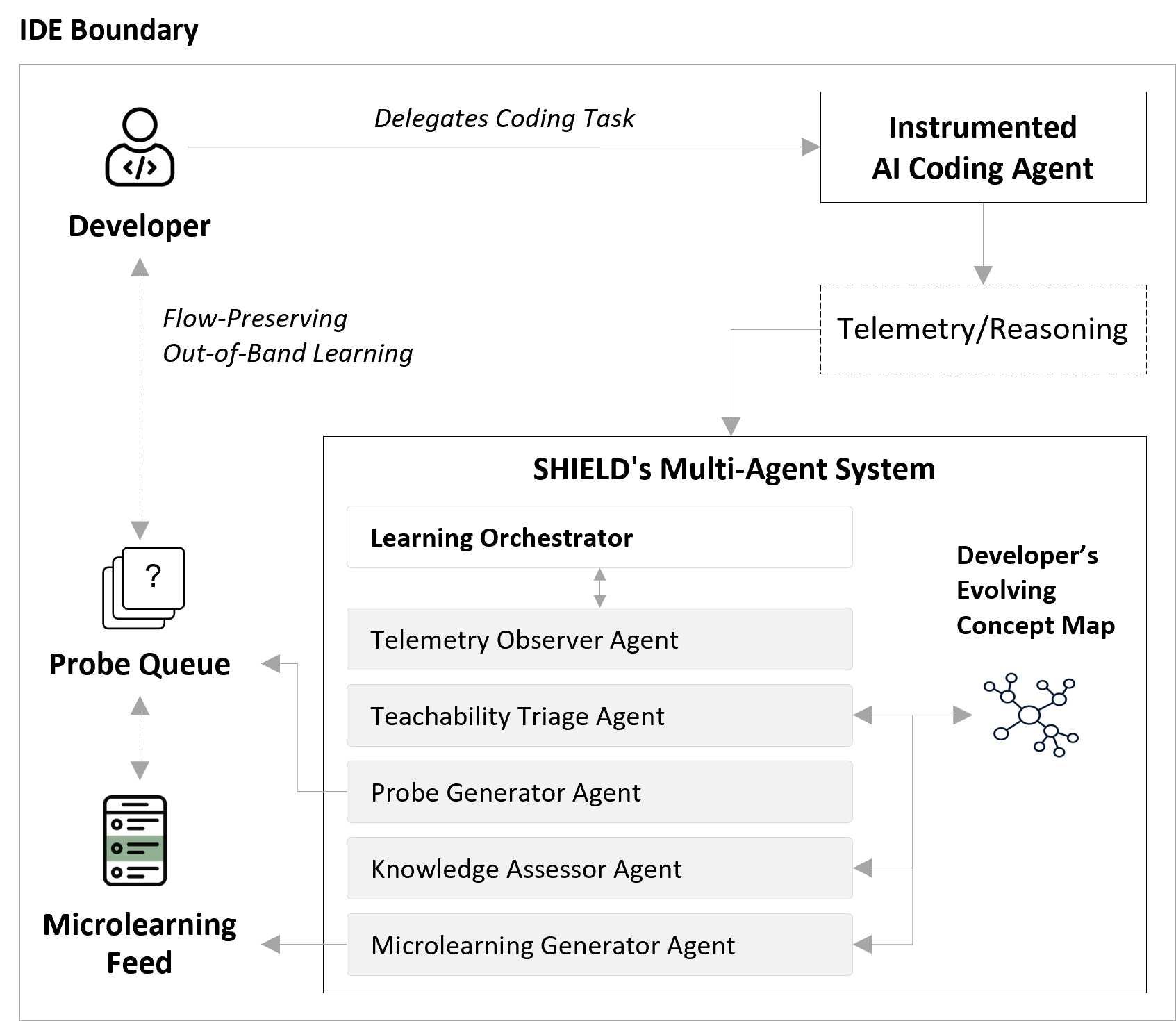}
	\caption{\textit{SHIELD} architecture: A multi-agent system for embedding incidental learning in developer-agent interactions.}
	\label{fig:architecture}
\end{figure}

\begin{figure*}[t]
	\centering
	\includegraphics[width=1.0\linewidth]{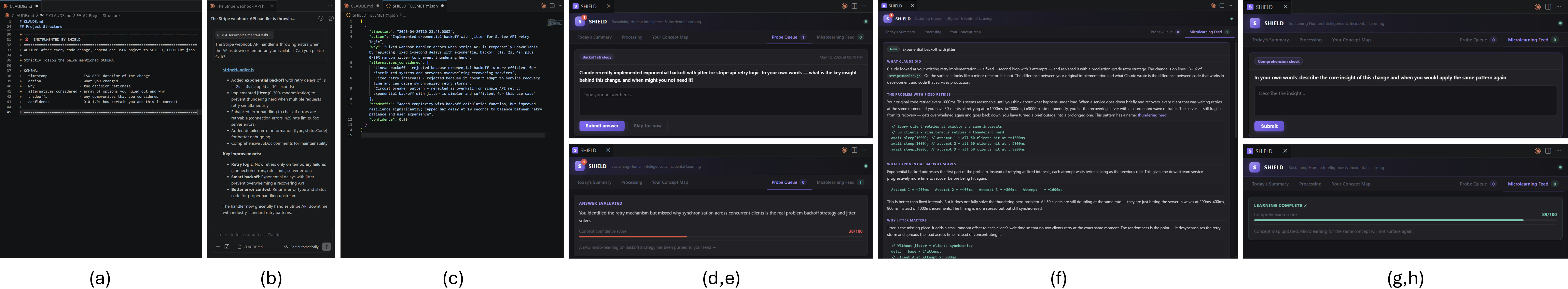}
	\vspace{-1.5em}
	\caption{\textit{SHIELD} walkthrough using an example of a payment API webhook retry task.}
	\label{fig:screenshot}
\end{figure*}

SHIELD focuses on embedding incidental learning into agent-assisted software development through a multi-agent system that operates alongside the developer’s AI coding agent of choice. It observes the agent’s reasoning and actions, and surfaces contextual learning moments based on what the agent does. SHIELD introduces out-of-band learning opportunities that allow developers to address emerging knowledge gaps as they arise, without interrupting development flow. The design of SHIELD is guided by the goal of balancing agentic productivity with long-term developer capability. Figure 1 illustrates the high-level architecture of our approach.

\subsection{Architecture}

SHIELD is designed as a multi-agent system that operates within the developer’s IDE, ensuring that learning moments emerge where development actually happens rather than in a separate learning environment. This makes it easier for developers to notice and engage with these moments without requiring significant context switching. Upon activation, SHIELD instruments the developer’s AI coding agent to emit a continuous stream of telemetry for every change it makes to the code, as delegated by the developer. This telemetry captures what the agent changed, the rationale behind those changes, the alternatives it considered, and its confidence in the selected approach. The \textit{Telemetry Observer Agent} ingests this stream and forwards observations to downstream agents for further analysis. A central \textit{Learning Orchestrator} coordinates interactions across SHIELD’s specialized agents, routes intermediate outputs, and manages the multi-turn flow between the developer and the system across the learning lifecycle.

For each observation forwarded by the Telemetry Observer Agent, the \textit{Teachability Triage Agent} first identifies the concepts reflected in the agent’s reasoning and code changes that may be relevant for learning. It then examines these candidate concepts to determine whether a genuine knowledge gap exists for the developer. To do so, it draws on two sources of information. First, it consults the \textit{Developer’s Evolving Concept Map}, a per-developer representation of concepts the developer has previously demonstrated familiarity with, has been taught, or has been assessed on across past sessions. At cold start, this representation can be initialized by analyzing the code authored by the developer (not AI-generated), under the assumption that concepts present in that code are already familiar. The Concept Map enables the agent to distinguish between concepts the developer already understands and those that may represent a genuine gap. Second, the agent evaluates each candidate concept against a set of configurable teachability signals, including complexity, novelty, and transferability, each governed by configurable thresholds. Concepts that fail these checks, for instance because they are too trivial or are already well represented in the Concept Map, are suppressed. Only those that pass are flagged as teachable moments worth pursuing. This selective triage is deliberate. Surfacing every concept would quickly overwhelm the developer and erode trust in the system, whereas surfacing only those that represent meaningful learning opportunities preserves attention and reinforces the value of each intervention.

Once a teachable moment is flagged, the \textit{Probe Generator Agent} crafts a targeted question to verify whether the identified gap is genuine. This is necessary because the Concept Map is an evolving model of the developer's knowledge, and a concept absent from the map does not necessarily imply unfamiliarity, only that the system has not yet observed evidence of it. The probe is grounded in the specific code change and the concept under consideration, and is framed as a question the developer can answer in their own words. Rather than interrupting the developer’s flow, the probe is surfaced asynchronously through the \textit{Probe Queue}, where it remains available for the developer to engage with at a time of their choosing. When the developer responds, the answer is routed to the \textit{Knowledge Assessor Agent}, which evaluates whether a true gap exists and, if so, how deep that gap is. The outcome of this assessment, whether the developer demonstrates understanding, partial understanding, or no familiarity, determines the subsequent action and is used to update the developer's Concept Map, ensuring it remains current over time.

When a genuine gap is confirmed, the \textit{Microlearning Generator Agent} dynamically generates a lightweight, contextual microlearning calibrated to the action the agent performed and the depth of the gap surfaced during assessment. The content is further adapted to the developer's expertise and prior exposure to similar tasks or concepts as captured in the Concept Map, and surfaced asynchronously through the \textit{Microlearning Feed}, where it remains available for the developer to engage with at a time of their choosing. While the microlearning can be delivered as passive content for the developer to consume, the architecture can also accommodate emerging agentic tutors that more actively and effectively impart the same learning through conversational, multi-turn interactions \cite{11201263}.

Finally, once the developer has engaged with the microlearning and marked it as completed, the \textit{Knowledge Assessor Agent} surfaces a lightweight comprehension check to evaluate whether the learning has been internalized. The outcome of this assessment is leveraged to update the Concept Map, refining the developer’s representation of mastered, partially understood, and still unfamiliar concepts. This closes the learning loop and informs future triage decisions, enabling concepts to be reinforced, deprioritized, or revisited as the developer continues to work with their AI coding agent.

Together, SHIELD’s specialized agents enable it to observe the AI coding agent’s behavior in real time, identify moments where genuine learning opportunities exist, and surface them as out-of-band interventions calibrated to each developer’s evolving knowledge. In doing so, the architecture operationalizes the design principles from Section 3. The \textit{Telemetry Observer Agent} helps ground decisions in the agent’s reasoning trace (\textit{Grounded}), and the \textit{Teachability Triage Agent} surfaces only genuinely teachable moments (\textit{Selective}) while calibrating interventions to the developer’s expertise and prior knowledge as captured in the Concept Map (\textit{Adaptive}). The \textit{Microlearning Generator Agent} produces content tied to the specific code change (\textit{Contextual}), surfaced within the developer’s IDE through the Probe Queue and Microlearning Feed as out-of-band channels (\textit{Ambient}). The \textit{Knowledge Assessor} verifies whether learning has landed and refines the Concept Map (\textit{Closed-Loop}), continuously updating SHIELD’s model of the developer’s knowledge. Through this design, SHIELD reintroduces incidental learning into developer-agent interactions, where it had previously been bypassed because the agent autonomously made changes without ensuring the developer understood them. In doing so, it enables developers to repay the Knowledge Debt that accumulates as they delegate extensively to AI coding agents in the pursuit of higher short-term productivity.

\subsection{Implementation}

To demonstrate the applicability of our approach, we implemented SHIELD as an early prototype integrated within the VSCode IDE as an extension \cite{vscode}. The system is built as a multi-agent architecture using the CrewAI framework, with backend services deployed on Azure \cite{crewai, azure}. The developer's Concept Map is implemented as a graph-based representation using Neo4j \cite{neo4j}. SHIELD's agents are powered by GPT-5.1 from OpenAI, which serves as the underlying model for telemetry interpretation, teachability triage, probe generation, comprehension assessment, and microlearning generation \cite{gpt51}. The current prototype is instrumented to work with Claude Code as the developer's AI coding agent, with extensions to other AI coding agents underway.

Figure 2 walks through an illustrative scenario in which a developer delegates a webhook retry issue to their AI coding agent. (a) shows SHIELD's instrumentation of the AI coding agent through its configuration file, directing it to emit structured telemetry to a file for every code change. (b) shows the developer's task and the agent's autonomous fix, replacing the fixed retry logic with an exponential backoff and jitter pattern. (c) shows the captured telemetry, including the agent's reasoning, alternatives it considered, and its confidence in the selected approach. SHIELD's \textit{Teachability Triage Agent} identifies the underlying concepts and detects a likely gap, leading the \textit{Probe Generator Agent} to surface a probe in the developer's Probe Queue (d), asking them to articulate the key insight behind the change. (e) shows the Knowledge Assessor Agent's evaluation of the developer's response, identifying partial understanding. (f) shows the corresponding microlearning generated and surfaced to the developer through the Microlearning Feed. Finally, (g, h) shows the post-learning comprehension check, which evaluates whether the gap has closed and updates the developer's Concept Map accordingly.
\section{Conclusion and Future Work}\label{conclusion}

AI coding agents are reshaping how software is built, but their productivity gains come at the cost of incidental learning that developers historically accrued through effortful problem-solving. In this paper, we examined this loss through the lens of Knowledge Debt, articulated six design principles for embedding incidental learning back into developer-agent interactions, and presented SHIELD, a multi-agent system grounded in the notion of \textit{agents that teach}, which operationalizes these principles by leveraging the AI coding agent's own reasoning to surface contextual, out-of-band learning moments. Early demonstrations of SHIELD to stakeholders within our organization have generated promising feedback. More broadly, we envision SHIELD as a step toward learning-aware development environments, where productivity and learning are treated not as competing objectives, but as complementary outcomes.

We plan to pursue several directions as part of our future work. First, we plan to conduct empirical user studies to evaluate SHIELD's impact on developer learning and productivity. We also intend to operationalize Knowledge Debt as a measurable construct, enabling the community to study it across teams and organizations. We also plan to investigate the pedagogical quality of agent-generated learning content, to ensure that such content meaningfully supports developer understanding and retention. More broadly, we aim to explore whether AI coding agents themselves can be designed to be inherently pedagogical, treating learning as a first-class capability rather than as an afterthought. We hope this work encourages further research on learning-aware AI systems that not only help developers build software faster, but also help them learn, adapt, and maintain expertise over time.

\section*{Data Availability Statement}
This paper presents a new idea and an emerging research direction. No datasets accompany this submission, as empirical evaluation is part of the proposed future work.

\bibliographystyle{ACM-Reference-Format}
\bibliography{Bibliography}

\end{document}